\documentclass[twocolumn,showpacs,preprintnumbers,amsmath,amssymb]{revtex4}

\usepackage{graphicx}
\usepackage{amssymb}
\usepackage{epstopdf}
\usepackage{float}      
\newcommand{\ignore}[1]{}
\newcommand{\be}{\begin{equation}} \newcommand{\ee}{\end{equation}}
\newcommand{\ba}{\begin{eqnarray}} \newcommand{\ea}{\end{eqnarray}}
\newcommand{\nn}{\nonumber} \renewcommand{\bf}{bf}
\newcommand{\ra}{\rightarrow}

\renewcommand{\a}{\alpha}

\newcommand{\sigmav}{{\langle \sigma v \rangle} }

\def\slashb#1{\setbox0=\hbox{$#1$}#1\hskip-\wd0\dimen0=5pt\advance
        \dimen0 by-\ht0\advance\dimen0 by\dp0\lower0.5\dimen0\hbox
          to\wd0{\hss\sl/\/\hss}}

\input{epsf}

\begin{document}

\title{ Relic Abundance Predicts Universal \\ {\it Mass-Width} Relations for Dark Matter Interactions}

\author{ Mihailo Backovi\'{c} and John P. Ralston}
  \affiliation{Department of Physics \& Astronomy, \\ The University of Kansas,
  Lawrence, KS 66045}

\begin{abstract}

We find new and universal relations for the properties of dark matter particles consistent with standard relic abundances. Analysis is based on
first characterizing the $s$-channel resonant annihilation process in great detail, keeping track of all velocity-dependence, the presence of
multiple scales and treating each physical regime above, below, and close to thresholds separately.  The resonant regime as well as extension to
include non-resonant processes are then reduced to analytic formulas and inequalities that describe the full range of multi-dimensional numerical
work. These results eliminate the need to recompute relic abundance model by model, and reduce calculations to verifying certain scale and
parameter combinations are consistent. Remarkably simple formulas describe the relation between the total width of an $s$-channel intermediate
particle, the masses and the couplings involved. Eliminating the width in terms of the mass produces new consistency relations between dark matter
masses and the intermediate masses. The formulas are general enough to test directly whether new particles can be identified as dark matter.
Resonance mass and total width are quantities directly observable at accelerators such as the LHC, and will be sufficient to establish whether new
discoveries are consistent with the cosmological bounds on dark matter.

\end{abstract}

\pacs{95.35.+d}

\maketitle

\section*{}

Thermal evolution of dark matter in the early universe can be found in textbooks to predict a velocity-averaged annihilation cross section
$\sigmav \sim 3 \times 10^{-26}cm^{3}/s$. However the textbook exercise happens to assume velocity-independent cross sections that are neither
general, nor reliable.  In this paper we give a complete and general analysis based on a new strategy. Our method solves the inverse problem of
bounding the multi-dimensional parameter regions such that the relic abundance is fixed. The analysis involves several novel steps.

Many models are summarized by $t$ and $u$-channel exchanges that are slowly varying, plus $s$-channel resonances that are a great complication. An
early study by Greist and Seckel  \cite{griest} noted that resonant processes violate the assumptions of constant cross sections, while being
impossible to summarize with equivalently simple formulas. A typical resonant calculation involves several coupling constants and 5 dimensionful
scales: the incoming energy, two masses $m_{X}$, $m_{Y}$, the final state mass or masses, and the final relic density. Choosing one point in this
huge parameter space and solving the Boltzmann relic evolution will predict one particular relic density. Complete exploration has previously
appeared impossible, and most studies are limited to checking consistency in a plane of a few selected parameters.

Our approach first characterizes relic evolution via $s$-channel annihilation. The work is partly numerical, and partly analytic; we make no
assumptions about masses, couplings, or final states. We find several tricks to identify the important scales and ratios of scales that describe
every possible parameter region. We use unitarity and the optical theorem to represent exactly the $s$-channel decay into all possible final
states.

As a result, we find universal mass-width relations which fit the numerical work across the whole parameter range. Let $X$ be the dark matter
(mass $m_{X})$, and $Y$ be the $s$-channel intermediate connector (mass $m_{Y}$, width $\Gamma_{Y})$, by which $X+X \ra Y \ra anything$. The
mass-v-width relation for the pole below threshold is \ba
     \Gamma_Y (m_Y < 2m_X)  &=&  {8 \over \pi}{1\, GeV \over C_{jj'} \alpha_{XXY}}   \left({ m_X \over 730 \, GeV} \right)^3 \nn \\
                            && \times \left(1 + {m_X \over 2 m_Y} \right) \left(1 - \frac{m_Y}{2 m_X} \right)^2.\nn  \label{widthmassrrel}
\ea Here $\alpha_{XXY}$ is the coupling of dark matter to the intermediate connector particle, and $C_{jj'}$ is a spin-counting coefficient from
which dimensionful scales have been removed. An equivalent formula for a pole above threshold ($m_Y > 2m_X$) is presented in Eq.
\ref{abovethreshold}, Section \ref{sec:above}.

To show how the relation works, suppose an $s$-channel connector of mass $m_{Y}=600$ GeV has coupling $\a_{XYY}=10^{-2}$ and width $\Gamma_{Y}=5$
GeV. Then $m_{X} =417.8$ GeV is the mass of the dark matter if the pole lies below threshold and dominates the relic evolution. This is far more
precise and specific than the order-of-magnitude estimates generated by $\sigmav  \sim  3 \times 10^{-26}cm^{3}/s$. If we have the same mass
$m_{X}$, and the width $\Gamma_{Y} = 2$ GeV, then the resonance $Y$ is too narrow to give the usual relic abundance.

Many examples of dominant $s$-channel dark matter annihilation models can be found in the literature \cite{Ibe:2009dx,Kadota:2010xm}. However we
are not limited to $s$-channel dominance. The most general cross section, including any number of channels and interference, is either larger or
smaller than the $s$-channel annihilation. Supposing that the cross section with extra non-resonant channels is larger provides an inequality of a
particular sense, given in Section \ref{sec:morechannels}. An inequality of the opposite sense comes from the opposite assumption.  In general the
$s$-channel dominant case produces a bounding surface in the space of all the parameters. The surface is simple enough that a number of powerful
inequalities in selected parameter planes come out rather easily. But we can do more: Eq. \ref{newanappx} shows quantitatively how to take into
account any amount of resonant versus non-resonant cross sections with a simple ``replacement rule.''

The mass-width relations also serve as a test of any new physics compared to dark matter cosmology. If the width and mass of a resonance do not
match our relations then it is not a candidate to produce relics. Conversely a match of mass and width would be an indisputable signal of
discovery. Note the width of a new particle is always {\it a physical observable} available from its production, allowing direct data-versus-data
comparisons not depending on model details. Thus we expect our formulas to be useful for dark matter studies at the LHC and the ILC
\cite{Cheung:2007ut,Kumar:2006gm,Han:2007ae}.

Moreover, {\it widths are calculable} in almost every model. The ansatz $\Gamma_{Y} =\a_{\Gamma}m_{Y}$ is typical, but we also use it as a {\it
definition} of the symbol $\a_{\Gamma}$ for analysis. Recall that the Standard Model $Z$-boson has a mass of $m_Z =91.2 GeV$ and a total width
$\Gamma_Z = 2.1 GeV$, giving an effective coupling of $\alpha_Z = 0.03$. The effective coupling absorbs all the channels, including the invisible
ones, all the couplings, spin factors, and phase space. Combining $\Gamma_{Y} =\a_{\Gamma}m_{Y}$ with Eq. \ref{widthmassrrel} yields a non-linear
equation relating $m_{X}$, $m_{Y}$ and the couplings. For $m_{X} >> 100$ GeV the relation reduces to $m_{Y} = 2 m_{X} + c_{1}+ c_{2}/m_{X}+...$
where $c_{1}$, $c_{2}$ are known functions of couplings. It is very surprising there is {\it always an allowed solution} for arbitrarily large
$m_{X}$, $m_{Y}$, contrary to the expectations of Born-level estimates. In the event that symbol $\a_{\Gamma}$ contains some mass dependence, as
with certain theories with dimensionful couplings, the mass relation remains good and can be explored without needing to repeat the relic
abundance calculations.

We built on previous experience with $s$-channel annihilation effects in the galactic halo \cite{Backovic:2009rw}. Velocity dependence in
annihilation came to the front with recent satellite data from PAMELA \cite{Adriani:2008zr}, FERMI \cite{Baltz:2008wd}, PPB-BETS
\cite{Torii:2008xu} and other experiments \cite{Barwick, chang}. Whether or not the data might be a signal, the studies have led to recognition
that Born-level cross sections are not adequate. Exaggerated claims about ``Sommerfeld factors,'' sometimes thought to be exact non-perturbative
effects, both violate general principles \cite{Backovic:2009rw} and phenomenological tests \cite{Dent:2009bv,Zavala:2009mi,Feng:2010zp}. Refs.
\cite{haloIbe, haloGuo, Backovic:2009rw} found the enhancements of ordinary resonant physics can be surprisingly large. Resonant processes can
saturate unitarity bounds on annihilation in halo circumstances, generating large ``boost factors'' suggested by the data. It is because certain
width-dependent effects ,perturbatively small in the high energy limit, may dominate everything in the non-relativistic limit. The problems of
hevy thremal relic evolution in the early universe are similar, because the abundance is determined primarily in a regime sensitive to
non-relativistic dynamics. Refs. \cite{Baro:2007em, Drees:2009gt} show concrete examples in which even the effects of one loop corrections on the
relic abundance can be significant.

There are many applications of our results. Section \ref{sec:cs} reviews the standard relic abundance formalism for completeness. Section
\ref{sec:schannel} presents the $s$-channel mass-width relation for a pole below threshold, along with motivation for the analytic formula. At
each stage where it is appropriate we convert the equalities developed for $s$-channel dominance into inequalities, illustrated with graphics.
Section \ref{sec:wid} explores consequences of calculable widths. This leads to $m_{X}-m_{Y}$ relations between the intermediate particle and the
dark matter. The surprising fact that $m_{Y}\sim 2 m_{X}$ can fit relic abundance for arbitrarily large masses is something to explore. It turns
out to be a generalization of the ``funnel' region'' known in minimal supersymmetric models (SUSY) \cite{Bhattacharya:2009ij, Djouadi:2005dz,
Ellis:2007by, Lahanas:2001yr, Herrmann:2007ku, Belanger:2006qa}, without making an assumption of SUSY.  We suspect that our relations are more
restrictive than the commonly known ones, and may rule out some models. Separate analysis for the case where the intermediate connector mass
$m_{Y}$ is above the kinematic threshold $2 m_{X}$ is given in Section \ref{sec:above}.

\section{Cross Sections and Relic Abundance \label{sec:cs} }

\begin{figure}[htb]
\begin{center}
\includegraphics[width=3.5in]{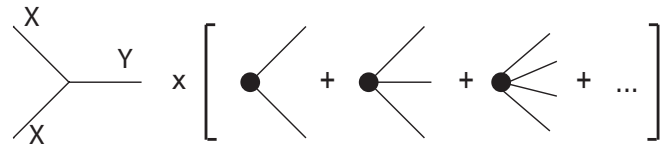}
\caption{ $s$-channel dark matter annihilation diagrams into all possible final states. \label{fig:diagram}}
\end{center}
\end{figure}

The thermally evolving number density  $Y$ is calculated as a function of the inverse temperature $x \equiv m_{X}/T$ using \ba \frac{dY}{dx}
=\xi(x, \, m_{X}) \sigmav(x) (Y^2 - Y^2_{EQ}), \nn \ea where $Y_{EQ}$ is the equilibrium density, and $ \xi(x,\, m_{X}) = - x s(x) / H(m_{X}) $ is
a combination of standard entropy and Hubble functions \cite{Kolb}. The asymptotic solution follows in the regime of $Y_{EQ} \ra 0$:  \ba
\frac{dY}{Y^{2}} &=&
\xi(x, \, m_{X})\sigmav(x)  dx; \nn \\
Y_\infty^{-1}  &=&  \int_{x_d}^\infty dx \,  \xi(x, \, m_{X}) \sigmav  . \label{xd}   \ea The lower limit $x_{d}$ is computed self-consistently.
Given the standard value of the critical density, the asymptotic density $\Omega = m_{X} Y_{\infty}s_{0}/\rho_{crit}$ follows, using the critical
density $\rho_{crit}$ and entropy density $s_{0}$ of the present universe.

Particle physics enters in the annihilation cross section $\sigma(v)$ and its thermal average $\sigmav (x)$. The rapid energy dependence and
multiple scales of resonant cross sections require special analysis. Expansions of the form $\sigmav \approx a + b/x + c/x^2 + ...$ are never
good approximations over the entire range of $x$.

Let $M_{XX \ra f}$ be the amplitude for $XX$ to go to a final state $f$. The cross section $\sigma$ goes like the amplitude-squared, summed over
all final states  (Fig.\ref{fig:diagram}), and integrated over final state phase space LIPS: \ba d\sigma \sim {1 \over flux} \sum_{f} \, |M_{XX
\ra f}|^{2} \, dLIPS. \nn \ea The total cross section into all possible final states is given by the optical theorem: \ba \sigma_{tot} = -{ 1\over
2 k E_{CM}} Im(M(s, \, t=0)).\nn \ea Here $k$ is the momentum of either particle in the center of mass frame, and $M$ is the elastic scattering
amplitude. For a given total center of mass energy $E_{CM}$ and its square $s$, the forward propagators of intermediate states $Y$ go like
$(s-m_{Y}^{2}+i m_{Y} \Gamma)^{-1}$, where $\Gamma$ is the total width. Let $g_{XXY}^{2} t_{jj'} $ be the component of the elastic amplitude
containing the couplings of the initial/final states of spin $j$ to an $s$-channel particle of spin $j'$. Then channel by channel, the optical
theorem
predicts \ba \sigma_{tot} & =& -{1 \over 2 k E_{CM}} Im \left({  g_{XXY}^{2} t_{jj'} \over s-m_{Y}^{2}+i m_{Y} \Gamma_{Y} }\right) \nn \\
& =& { g_{XXY}^{2} \over 2 k E_{CM}} {m_Y \Gamma_{Y} t_{jj'} \over ( s-m_{Y}^{2}  )^{2}+ m_{Y}^{2}\Gamma_{Y}^{2}  }. \label{relbw} \ea The symbol
$\Gamma_Y$ represents the total width of $Y$ to all final states, which allows us to describe numerous models with a single parameter.

In standard convention for amplitudes, the Feynman rules contain in/out state polarization and vertex factors compiled into the symbol $t_{jj'}$.
It is important to extract the mass ($m_{X})$ dependence of these factors for analysis. We define \be
    t_{jj'} = 4m_{X}^2 C_{jj'},
\ee whereby $C_{jj'}$ is typically a number of order unity. Scaling like $m_{X}^{2}$ is expected from dimensional analysis, and inevitable when
the initial state is dominated by the mass as the largest scale. We emphasize that $C_{jj'}$ is a definition that allows for any model while
postponing spin-sums and vertex factors until a model is chosen. For example, the annihilation of unpolarized Dirac Fermions via a $\gamma^{\mu}$
vertex produces $C_{{1 / 2} \, 1} =3/4.$ Another example is the annihilation of two scalar particles to a vector, which will have two derivatives
going like $k^{2} \sim m_{X}^{2}$, times the polarization sum over the vector particle modes. The coupling and spin factors then appear in the
combination $4 \pi a_{XXY}C_{jj'}$, where $\a_{XYY} = g_{XXY}^{2}/(4 \pi)$.

\begin{figure}[htb]
\begin{center}
\includegraphics[width=3in]{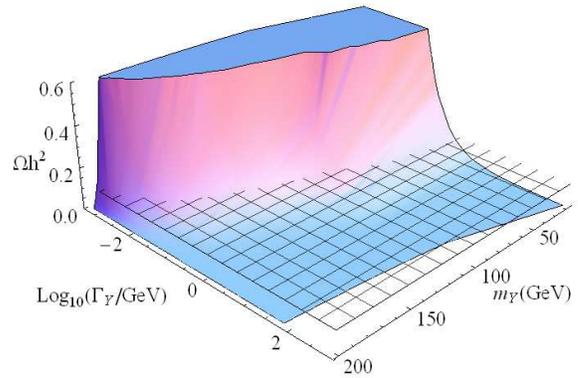}
\caption{Intersection of the mesh plane representing $\Omega h^2 = 0.1$ and the surface $\Omega h^2 (\Gamma_Y, m_Y, m_*, \alpha_*)$ gives a unique
curve $\Gamma(m_Y)$. $m_* = 100 GeV$, $\alpha_* = 0.01$ for the purpose of the graphic. } \label{fig:omegas}
\end{center}
\end{figure}

Now all $m_{X}$ dependence of the cross section, rate $\sigmav$, and density $Y_{\infty}$ dependence has been scaled out, except for a minor
cutoff ($x_{d}$) dependence in Eq. \ref{xd}, which must be reconciled numerically. In the numerical work we first choose a particular dark matter
mass $m_{X} = m_{*}$ and coupling $\a_{*}$, and then compute thousands of relic densities covering the $\Gamma_{Y}$, $m_{Y}$ plane. The condition
$\Omega (\Gamma_{Y}, \, \, m_{*}, \, m_{Y}, \a_* )h^{2}\ra 0.1$ produces a unique curve $\Gamma_{Y}$ versus $m_{Y}$, namely the function
$\Gamma_{Y}=\Gamma_{Y}(m_{Y}; \, m_{*}, \a_{*})$, as seen in Fig. \ref{fig:omegas}. With the scaling relations in hand, the curves are extended to
numerical predictions for the general functional dependence of $\Gamma= \Gamma(m_{Y}; \, m_{X}, \a_{XYY})$ consistent with a fixed relic density.

\section{Mass-Width Relations: Pole Below Threshold ($m_Y < 2m_X$) \label{sec:schannel}}

\begin{figure}[htb]
\begin{center}
\includegraphics[width=3in]{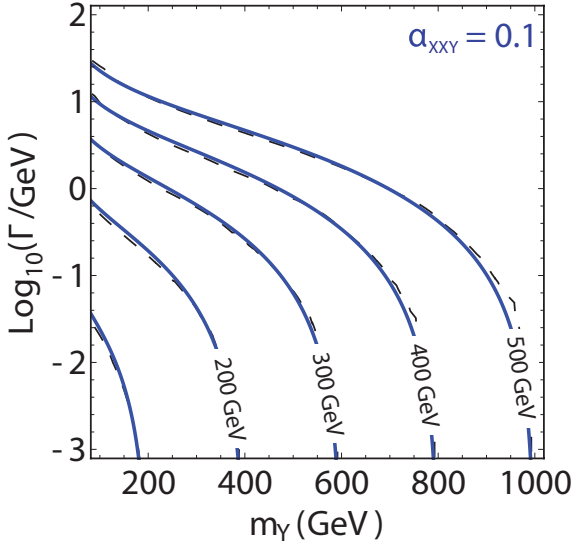}
\caption{ Relation of the $s$-channel width $\Gamma_{Y}$ and mass $m_{Y}$ {\it for a pole below threshold} consistent with cosmological relic
density $\Omega h^{2}=0.1$. Dashed curves (black online) are the numerical calculation. Solid curves (blue online) are the analytic relation of
Eq. \ref{anappx}. Each curve is evaluated with a fixed dark matter mass $m_{X}$ =100-500 GeV in 100 GeV increments.}
 \label{fig:GammaVMy.eps}
\end{center}
\end{figure}

Fig. \ref{fig:GammaVMy.eps} shows the mass-width relation as a family of curves plotted for selected $m_{X}$. The trend is that the further the
$m_{Y}$ is from the threshold, the larger $\Gamma_{Y}$ must be needed to keep relic densities constant, and vice versa. This is because the
proximity to threshold (rather than the absolute size) of $m_{Y}$ is the dominant effect. Poles closer to the threshold make for larger cross
sections, which need to be compensated by smaller width.

This quantitative understanding leads to remarkable analytic formulas reproducing the whole parameter range.

\subsection{Analytic Representation}

Observe in Fig. \ref{fig:GammaVMy.eps} that each dashed (black) curve moving to the right terminates in a region of $\Gamma_{Y}\ra 0$. Near the
threshold everything is determined by the degree of the zero of the function, $\Gamma_Y \sim (m_{Y}- 2 m_{X} )^{n}$. The power $n = 2$ can be
gotten analytically, but was also fit directly with numerical work. Then we know $\Gamma_{Y} \sim (1 -m_{Y}/2m_{X})^{2}$, times a known factor of
$m_{X}$

In the opposite extreme of $m_{Y} << 2 m_X$  the velocity averaged cross section $\sigmav$ reduces to another simple analytic result: \be
    \sigmav (m_{Y} << 2 m_X) \ra \frac{\pi \alpha_{XXY} C_{jj'}}{4 m_X^4} m_Y \Gamma_Y \label{smallmysig}
\ee Notice the dependence going like $1/(m_{Y}\Gamma_{Y})$. Inverting the equation gives \be
    \Gamma_{Y}  \ra {4 m_X^4 \over \pi \alpha_{XXY} C_{jj'} m_Y} \sigmav \ee

The limit of small connector mass $m_{Y} \ra 0$ with $\Gamma_Y << 2m_X $ approaches the Born approximation, and for us is the unique case where
the Born cross section is relevant. In the Born limit we know $\sigmav  \ra 3 \times 10^{-26} cm^3/s$, which fixes one overall scale. Then
accounting for factors of $m_{X}$ gives \ba   \Gamma_Y   \sim    {4 m_{X}^3 \over \pi C_{jj'} \alpha_{XXY}} (2.6 \times 10^{-9} GeV^{-2})  \nn \\
\times  \left(1 - \frac{m_Y}{2 m_X} \right)^2 { m_{X} \over m_{Y}} g(m_{Y}/m_{X}). \nn \ea The dimensionless interpolating function
$g(m_{Y}/m_{X})$ remains. It must obey $ g \ra 1$ when $m_{Y}<< 2m_{X}$, suggesting a polynomial expansion $g \sim 1+ \sum_{k}
\,g_{k}(m_{Y}/m_{X})^{k}$. Two terms suffice with $g_1 = 2$. Our analytic formula for the {\it pole below threshold mass-width
relation} is then \ba
   \Gamma_Y &=&   {8 \over \pi} {1\, GeV \over C_{jj'} \alpha_{XXY}}    \left({ m_X \over 730 \, GeV}\right)^3 \nn \\ & \times  &\left(1 + {m_X \over 2 m_Y} \right) \left(1 - \frac{m_Y}{2 m_X} \right)^2,  \nn \\  & \: & \: \:\: \:\:  for  \: m_{Y}< 2 m_{X}   \label{anappx}
\ea  The fit of the formula to the numerical work is extremely good. Fig. \ref{fig:GammaVMy.eps} shows a typical example.

We have already noted that Eq. \ref{anappx} generalizes and replaces the traditional formula $\sigmav = 3 \times 10^{-26} cm^3/s$. The formula
accounts for the fact that intermediate states of all particles coupling to dark matter are either absolutely stable or have a finite lifetime. To
underscore the difference, compare the traditional counting rules, motivated on dimensional analysis, using annihilation cross sections of order
1-picobarn. Under the assumption $\sigma v \sim \a_{X}^{2}/m_{X}^{2} \sim pb$, a typical upper limit $\a_{X}^{2} \lesssim 10^{-4}$ would imply
$m_{X} \lesssim 200$ GeV. This well-known result needs a finely tuned mass-couping relation to make a Universe. Our formula contains far more
information, and reveals the hidden assumption that $m_{Y}<<m_{X}$ was implicitly assumed for the traditional formula to be consistent.

\subsection{Replacement Rule for Adding Non-Resonant Channels}

\label{sec:morechannels}

Some models of dark matter annihilation include more than one channel. No matter how many channels are involved, as long as the $s$-channel is a
part of the model, the mass-width relation of Eq. \ref{anappx} can be generalized.

\begin{figure}[htb]
\begin{center}
\includegraphics[width=3in]{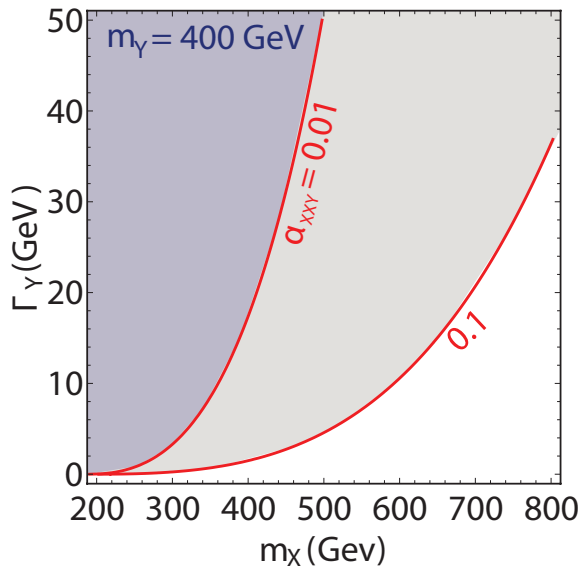}
\caption{Upper limits on $ \Gamma_Y $ assuming $s$-channel annihilation (pole below
threshold) plus other channels increasing the cross section. Shaded regions to the left and above the contours are not allowed. Curves show different couplings $\alpha_{XXY} = g_{XXY}^2/4\pi$; $m_Y =400$ GeV is used for the purpose of the graphic. Larger $m_Y$ pushes contours to the right. \label{fig:mxagplot}}
\end{center}
\end{figure}

For definiteness, suppose the addition of other processes {\it increases} the annihilation rate. Then the theory keeping $\Omega h^2$ fixed will
require smaller  $\Gamma_Y$, all other things fixed. That condition rules out all contours {\it to the right and above the contours} shown in Fig.
\ref{fig:GammaVMy.eps} created by $s$-channel dominance. The {\it allowed region to the right and below} each line implies an inequality (Fig.
\ref{fig:mxagplot}) : \ba
   \Gamma_Y &\leq&   {8 \over \pi}\left({1\, GeV \over C_{jj'} \alpha_{XXY}}\right)    \left({ m_X \over 730 \, GeV}\right)^3  \nn \\
   && \times\left(1 + {m_X \over 2 m_Y} \right) \left(1 - \frac{m_Y}{2 m_X} \right)^2 . \label{inequality} \ea  While cross sections often increase when
channels are added, destructive interference occurs in some models. In that case the inequality reverses the sign.

To illustrate the use of Eq. \ref{inequality}, suppose a new vector boson ($Z'$ perhaps) is discovered at the LHC. Measuring the resonance observed in \textit{any} channel will give its mass $m_Y$, and the \textit{total} width $\Gamma_Y$.  Applying Eq. \ref{inequality} then
gives the consistent $m_X, \, \alpha_{XXY}$ parameter space regions consistent with relic abundance \textit{without a need for an extensive numerical parameter space scan}. These {\it predicted} coupling relations are then compared to the information from production rate and branching ratio into particular channels seen in the experiment.

A new relation comes from a ``replacement rule.'' Let the total velocity averaged cross section be expressed as \be
    \sigmav_{tot} = \sigmav_s + \sigmav_{other} .\nn
\ee Suppose $\sigmav_{other}$ happens to be consistent with the traditional Born-style of approximation, by which \ba \sigmav_{other} = \sum_i
\alpha^i_{eff} / m_X^2 , \nn \ea where $\alpha^i_{eff}$ is an effective coupling to the $i^{th}$ channel. (An example model in the context of
heavy hidden sector dark matter can be found in Ref. \cite{MarchRussell:2008yu}.) Matching the extreme limits produces the replacement rule. If
the $s$-channel pole is near the threshold, the mass width relation should approach Eq. \ref{anappx}. If the pole is far from threshold the
resonant cross section approaches an effective Born-level cross section, and adds to it. That implies a boundary condition of $ \sigmav_s +
\sigmav_{other} \approx 10^{-9} GeV^{-2}$ at this endpoint. Reviewing how that scale previously entered the analysis suggests a replacement
rule:\be
    2.6 \times 10^{-9} GeV^{-2} \rightarrow  2.6 \times 10^{-9} {GeV}^{-2} - {\sum_i \alpha^i_{eff} \over m_X^2}
\ee

\begin{figure}[htb]
\begin{center}
\includegraphics[width=3in]{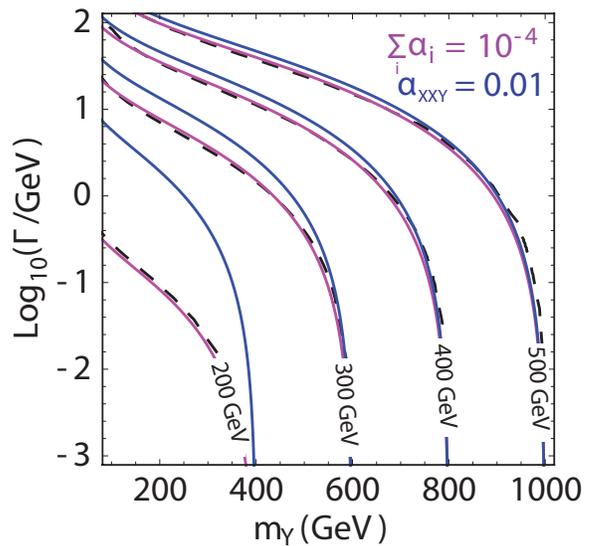}
\caption{Generalization of the mass-width relation to include ``Born-like'' channels. Black dashed curves show numerical evaluation. Solid curves
(magenta online) are the revised fit of Eq. \ref{newanappx}. Thick solid curves (blue online) are the approximations of Eq. \ref{anappx},
consistent with the role as an upper bound. Different curves use different masses $m_X = 100-500 GeV$, from left to right. Parameter $\alpha_{XXY}
= 10^{-2}$ and $\sum_i \alpha_{eff}^i = 10^{-4}$ for the purpose of the graphic. \label{fig:allchannels}}
\end{center}
\end{figure}

The revised mass-v-width relation then becomes  \ba \Gamma(m_Y) &=& {8 \over \pi}{1\, GeV \over C_{jj'}\alpha_{XXY}}
\left[ 1 -  \sum_i \alpha^i_{eff} \left({730 \, GeV \over m_X}\right)^2 \right] \nn \\
&& \times \,\, ({ m_X \over 730 \, GeV})^3 \left(1+\frac{m_X}{2m_Y} \right)\left( 1 - \frac{m_Y}{2m_X} \right)^2 \nn \\  \label{newanappx}\ea Fig.
\ref{fig:allchannels} shows the replacement rule performs quite well. Not surprisingly, the difference relative to a pure $s$-channel
annihilation model increases for smaller $m_X$. This is because the individual cross section contributions from other channels shown in Eq.
\ref{newanappx} scale uniformly like $1/m_X^2$.

\section{Calculable Widths Constrain the Masses \label{sec:wid}}

Up to here we have considered the width $\Gamma_Y$ as an independent parameter. In this section we go a step further and consider widths as
quantities which can be calculated. When we say that ``widths are calculable'' it emphasizes the facts that (1) most theories are perturbatively
coupled, and (2) most of the width will usually occur in a finite number of channels. {\it Whatever the model}, combining the calculation of the
width with the mass-width relation creates a new relation.

The differential rate $d\Gamma$ of a general decay of a particle of mass $m_{Y}$ is given by \ba d\Gamma \sim { 1 \over 2 m_{Y}  } |M|^{2}dLIPS.
\nn \ea Symbol $M$ is the amplitude. The final state phase space of two identical particles yield $\int dLIPS_{2} \sim  v_{f}$ where $v_{f}$ is
the velocity of either final state particle in the center of mass frame. In many cases the width is dominated by relativistic final states, $v_{f}
\ra 1$. It would be unusual, and a case of rather fine tuning, for all channels with phase space limitations $v_{f}<<1$ to dominate the total
width. Barring that event, by dimensional analysis, the width of a heavy particle with dimensionless coupling generally goes like its mass: \ba
\Gamma_{Y} \sim \a_{\Gamma} m_{Y}. \label{defn} \ea We make this an equality allowing symbol $ \a_{\Gamma}$ to absorb coupling constants, the
number of important channels, and model details. The general scaling of widths-proportional-to-mass is rather kinematic. However, if a
dimensionful coupling is introduced, then the mass dependence of {\it rest} of the calculation is dominated by dimensional analysis again. Keeping
in mind that $\a_{\Gamma}$ stands for the width actually calculated in a particular model, we continue.

\begin{figure}[htb]
\begin{center}
\includegraphics[width=3in]{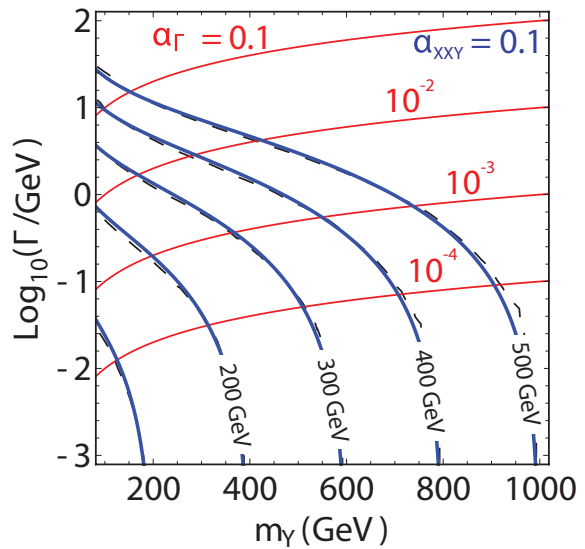}
\caption{ Combining the {\it below-threshold} width-v-mass $m_{Y}$ relation of Fig. \ref{fig:GammaVMy.eps} with $\Gamma_{Y} =\a_{\Gamma}m_{Y}$
represented by solid thin curves (red online). Intersections of the curves predict a non-linear relation between $m_{X}$ and $m_{Y}$ (text).
Values of $\a_{\Gamma} =10^{-1}$ (top curve) range to $\a_{\Gamma}=10^{-4}$ (bottom curve) in factor of 10 increments. }
 \label{fig:GammaVMyAdd.eps}
\end{center}
\end{figure}

With $\a_{\Gamma}$ fixed, the formula for $\Gamma_{Y}$ is an {\it increasing} function of $m_{Y}$. Meanwhile the $s$-channel mass-v-width
requirements are all {\it decreasing} functions of $m_{Y}$ (Fig. \ref{fig:GammaVMy.eps}, Eq. \ref{anappx}, Eq. \ref{newanappx}). Then the
width-v-mass relation always matches the $s$-channel mass-v-width relation at a definite point. Fig. \ref{fig:GammaVMyAdd.eps} shows $\Gamma_{Y} =
\a_{\Gamma} m_{Y}$ as red curves, whose intersections with the blue curves constrain the masses.

Inspection finds a surprising fact. Rather weakly coupled theories ($\alpha_\Gamma \lesssim 10^{-4}$) only intersect the relic curves in the
region where $m_Y \approx 2m_X$. In minimal supersymmetry, this result corresponds to the so called higgs "funnel" region of $m_0, m_{1/2}$
parameter space. However, our result is much more general and extends beyond the assumptions of SUSY.

The two couplings, $\a_\Gamma$ and $\a_{XXY}$ act in the same direction. Smaller $\a_{\Gamma}$ makes smaller widths that force the system into the
threshold region to be viable. Smaller $\a_{XYY}$ worsens the situation by pushing the contours of constant $\Omega h^2$ up in the
$\Gamma_{Y}-m_{Y}$ plane. The trend of both pushes masses into very near coincidence of $m_{Y} \sim 2 m_{X}$, which we call a ``finely-tuned
threshold.''

{\it Except for bound state formation} we have no reason to consider finely-tuned thresholds very plausible, but we can afford to stay neutral.
Bound states have been discussed in detail in Refs. \cite{Shepherd:2009sa, MarchRussell:2008tu}. Some basic relations between bound state widths
and masses are reviewed in Ref. \cite{Backovic:2009rw}. Bound state relations are very specific and require separate treatment that is not our
topic here.

 Our mass-width relation allows for classification of models according to the degree of fine
tuning. For example, Fig. \ref{fig:GammaVMyAdd.eps} shows a theory with $\alpha_{XXY} = 0.1$. The intersections are not very demanding, and the
theory is not finely tuned for $\alpha_\Gamma \geq 10^{-3}$. However the same theory using $\alpha_{XXY} = 10^{-3}$ will require widths $100$
times bigger for the same $m_Y$ to keep $\Omega h^2$ constant. At that point all contours are pushed up to such a degree we'd find the theory
finely-tuned for all reasonable $\Gamma_Y$.

\begin{figure}[htb]
\begin{center}
\includegraphics[width=3in]{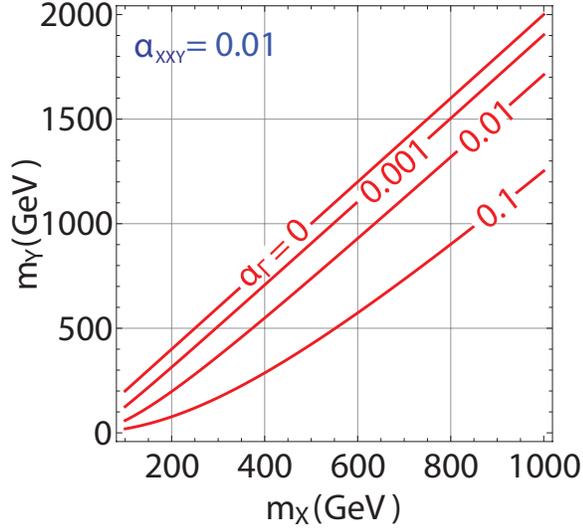}
\caption{ Mass of dark matter $m_{X}$ versus the mass of the particle in the $s$-channel $m_{Y}$. Red lines represent $\a_\Gamma = 0.001, 0.01, 0.1$ from top to
bottom. $\a_{XXY} = 0.01$ for the purpose of the graphic. Small widths ($\a_\Gamma$ small) require fine mass tuning, $m_Y \approx 2m_X$ to
accommodate correct relic abundance. \label{fig:mxmyplot}}
\end{center}
\end{figure}

\subsection{Dark Matter and Pole Mass Relations}

So far we have looked at an $m_Y$ - $\Gamma_Y$ relationship, given $m_X$. It is very interesting to consider the relation between $m_Y$ and $m_X$
given $\alpha_\Gamma$. Fig. \ref{fig:mxmyplot} shows the mass relationships for different values of $\alpha_\Gamma$. Once again $\a_{\Gamma}
\ra 0$ forces the resonance into the finely tuned region of $m_Y \approx 2m_X$.

Solving Eq. \ref{anappx} yields a cubic equation fixing $m_Y =m_Y(m_X, \alpha_{XXY}, \alpha_\Gamma)$. The relationship is nearly linear for a wide
range of $m_X $. Collect the couplings into a new symbol \ba \alpha_{\kappa}^{2} ={ \a_{XXY}\a_{\Gamma}C_{jj'} \over 10^{-4}} \nn \ea  Note the
symbol has been re-scaled in units of $ \a_{XXY}/10^{-2}$, $\a_{\Gamma}/10^{-2}$ we find reasonable. The series expansion for large $m_{X} $ is
found to be \ba   m_Y \approx 2m_X -313 GeV \alpha_{\kappa} +40.2 GeV\,  \alpha_{\kappa}^{2} \,  ({730 GeV \over m_{X}}) \nn \\ -  5.87 GeV \,
\alpha_{\kappa}^{3} \, ( {730 GeV \over m_{X}})^{2}+... \label{approx2} \ea Eq. \ref{approx2} with only the first two terms kept is essentially
exact for $\alpha_{\kappa} \lesssim 1, \, m_{X} \gtrsim 100$ GeV, while for $\alpha_{\kappa} \gtrsim 1$ a numerical evaluation is preferable.

\begin{figure}[htb]
\begin{center}
\includegraphics[width=3in]{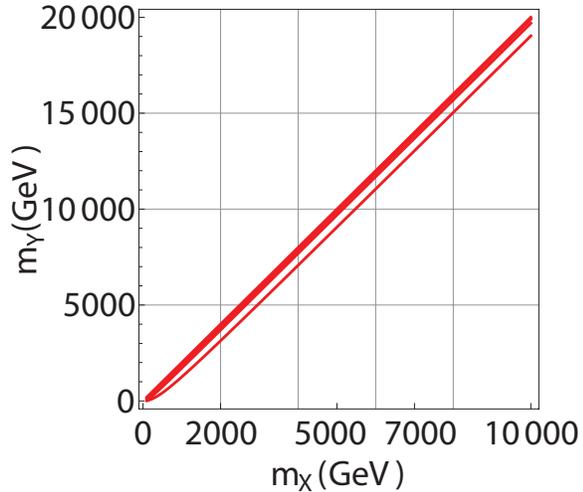}
\caption{ Same as Fig. \ref{fig:mxmyplot} but for an extended range of $m_X$. Red lines represent $\a_\Gamma = 0.001, 0.01, 0.1$ from top to
bottom. $\a_{XXY} = 0.01$ for the purpose of the graphic. \label{fig:mxmyplot2}}
\end{center}
\end{figure}

Analyzing Fig. \ref{fig:mxmyplot} and \ref{fig:mxmyplot2} we notice that the mass range of 100-500 GeV does not require extreme fine tuning for a
reasonable range of perturbative couplings. On the other hand the regime of $m_X >> 100 GeV$ seems to require a pole tuned very finely according
to Eq. \ref{approx2}.

\section{Mass-Width Relations: Pole Above Threshold ($m_Y > 2m_X$) \label{sec:above}}

\begin{figure}[htb]
\begin{center}
\includegraphics[width=3in]{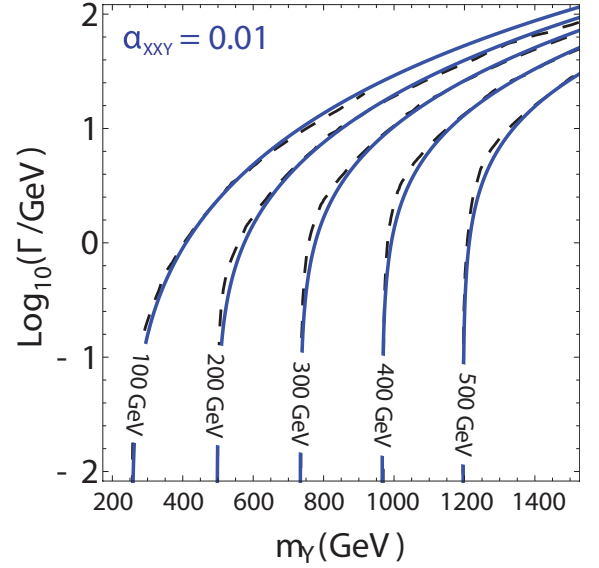}
\caption{ Relation of the $s$-channel width $\Gamma_{Y}$ and pole mass $m_{Y}$ {\it above threshold} consistent with cosmological relic density $\Omega
h^{2}=0.1$. Dashed curves (black online) are the numerical calculation. Solid curves (blue online) are the analytic relation of Eq.
\ref{abovethreshold}. Each curve is evaluated with a fixed dark matter mass $m_{X}$ =100-500 GeV in 100 GeV increments.}
 \label{fig:abovethreshold}
\end{center}
\end{figure}

The relic abundance calculation for a pole above threshold is complicated by a saddle point in the integration of $\sigmav$. To begin we again
consider the extreme limits. For $m_Y
>> 2m_X$ the velocity averaged cross section reduces to \be
    \sigmav(m_Y >> 2m_X) \rightarrow {4 \pi \alpha_{XXY} C_{jj'} \over m_Y^3} \Gamma_Y. \label{largemy}
\ee  By construction this limit reproduces the Born-level estimate, $\sigmav \approx 10^{-9} GeV^{-2}$. Introduce a dimensionless function $h$ to
describe other limits, expressed by \ba
    \Gamma_Y &\sim& {m_Y^3 \over 4\pi \a_{XXY} C_{jj'} } (6.4 \times 10^{-9} GeV^{-2}) \nn \\
            && \times h(m_Y, \, m_X, \,  A, \,  \alpha_{XXY}C_{jj'}). \nn
\ea We have normalized $h \ra 1$ for $m_Y >> 2m_X$ by absorbing the overall normalization into $6.4 \times 10^-9 GeV^{-2}$, suggesting the ansatz
\ba
    h = 1 - { \eta \, m_X \over m_Y} , \label{h}
 \ea where $\eta$ is the measure of the "offset" of $m_Y$ from the threshold $2m_X$.
Unlike the case of pole below threshold, the saddle point causes $\eta$ to be a function of $m_{X}$ and a relic scale parameter, which we call
$A$.

\begin{figure}[htb]
\begin{center}
\includegraphics[width=3in]{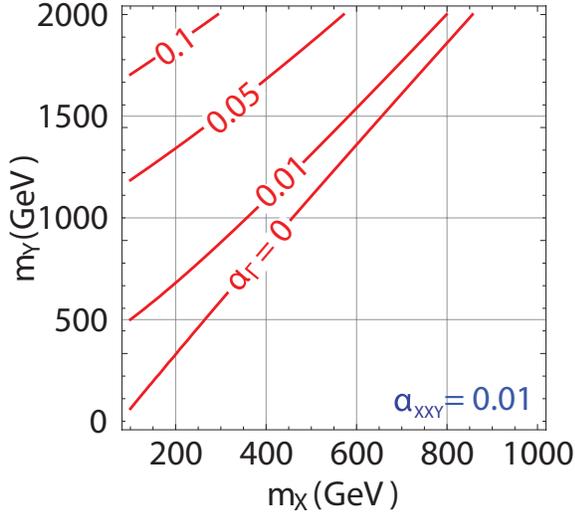}
\caption{Mass of dark matter $m_{X}$ vs. the mass of the $s$-channel particle $m_{Y}$ for a pole above threshold, $m_{Y}> 2 m_{X}$  Red lines represent $\a_\Gamma = 0, 0.01,
0.05, 0.1$ from top to bottom. $\a_{XXY} = 0.01$ for the purpose of the graphic. Small widths ($\a_\Gamma$ small) require very fine tuning of masses,
$m_Y \approx  \eta m_X$ to accommodate correct relic abundance. }
 \label{fig:mxmyabove}
\end{center}
\end{figure}

The extreme of $\Gamma \rightarrow 0$ gives more information about the function $h$. The Breit-Wigner factor can be approximated as \be
    {\Gamma m_Y \over (s-m_Y^2)^2 + (m_Y \Gamma)^2} \rightarrow \pi \delta(s - m_Y^2).
\ee The Boltzmann equation is then solved analytically in terms of error functions, predicting $h$ and $\eta$ in this limit: \be    \eta(m_X,
\alpha_{XXY}, C_{jj'}) \equiv 2 \sqrt{1 + {2 \,\over x_d}  erfc^{-1}({A \,  m_X^2 \over \alpha_{XXY} C_{jj'} }) },  \label{eta}\ee where $A = 1.3
\times 10^{-11} GeV^{-2}$ gives a good fit for all reasonable $m_Y$ and $\Gamma_Y$. The lower integration limit $x_d$ is computed in a self
consistent way and we find that the standard value of $x_d = 30$ is appropriate.

Eq. \ref{eta} involves the inverse complementary error functions ($erfc^{-1}$), which is somewhat cumbersome. While many numerical packages
(including Mathematica)  compute it, a simpler analytic formulation of $\eta$ is useful. Let $
    z =   A \, m_X^2 / \alpha_{XXY} C_{jj'}$. We find the approximation
\ba
    \eta(z) \approx 1.978 - 0.521 z - 0.051 Log[z] \nn
\ea
 is almost exact in the range $ 10^{-8} \leq z \leq 1$.

Our analytic formula for a pole above threshold is now: \ba
    \Gamma_Y &=& {1 GeV \over 4 \pi \alpha_{XXY} C_{jj'}}\left({m_Y \over 589 GeV}\right)^3 \left( 1 - {\eta(z) \, m_X \over m_Y}\right), \nn \\
    \nn \\  & \: & \: \:\: \:\:  for  \: m_{Y} > 2 m_{X}  \label{abovethreshold}
\ea Once again, the analytic approximation matches numerical work remarkably well. Fig. \ref{fig:abovethreshold} shows an example.

Eq. \ref{abovethreshold} reveals more finely tuned parameter regions for a pole above threshold. In the limit $\Gamma_Y \rightarrow 0$, $m_Y$ is
finely tuned to $\eta m_X$, as seen in Fig. \ref{fig:mxmyabove}. The competition between the pole position, width, and thermal Gaussian are all summarized by this generalization of the pole below-threshold relation.

As before, eliminating $\Gamma_{Y} =\a_{\Gamma}m_{Y}$ produces an $m_{X} - m_{Y}$ relation:  \ba  m_{Y} = {\eta(z) m_{X}\over 2} + {1 \over 2}
\sqrt{ \eta^{2}(z) m_{X}^{2}+  16 \pi \mu^{2} \a_{\Gamma}\a_{XXY}C_{jj'} } , \nn \ea where $\mu^{2} =589^3 GeV^{2}$.

\begin{figure}[htb]
\begin{center}
\includegraphics[width=3in]{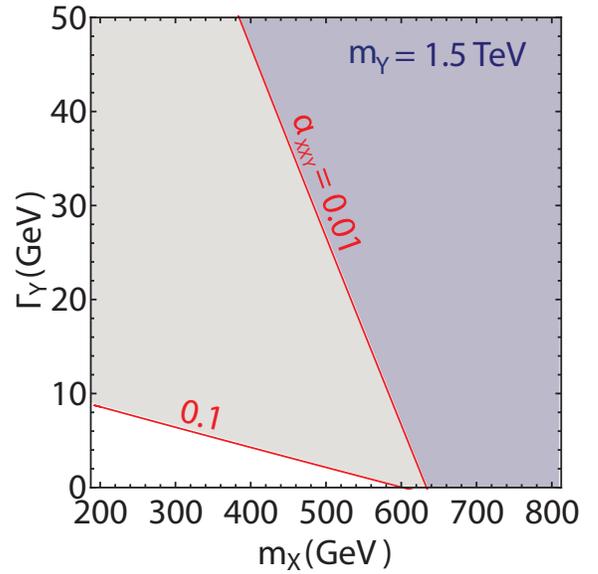}
\caption{Typical upper limits on $ \Gamma_Y$ given a mass $m_Y$ for different masses of dark matter $m_X$ assuming both $s$-channel annihilation (pole above
threshold) and other non-resonant channels. Shaded regions to the left and above the contours are not allowed. $m_Y =
1.5$ TeV was used for the purpose of the graphic; larger $m_Y$ pushes contours to the right. \label{fig:aboveineq}}
\end{center}
\end{figure}

\subsubsection{Upper Limit on $m_{X}$}

The particular form of the "offset function" $\eta$ yields an upper limit on $m_X$. From the derivation the argument of $erfc^{-1}$ must be less than one, which implies  \be
    m_X \leq 2.77 \times 10^5 GeV \sqrt{\alpha_{XXY} C_{jj'}}
\ee This formula is more precise than supposing $m_{Y}$ is bounded by a Born-level estimate and $2m_{X} <m_{Y}$. Consider for example a small coupling $\alpha_{XXY} = 10^{-4}$. Consistency with relic abundance requires dark matter masses $m_X \lesssim 2.8 TeV$.

\subsubsection{Inequalities for Non-Resonant Channels}

Generalization of the above-threshold mass width relation to allow for non-resonant channels is similar to the below-threshold case. When
$\sigmav_{tot} \geq \sigmav_s$ the mass width relation becomes \be
     \Gamma_Y \leq {1 GeV \over 4 \pi \alpha_{XXY} C_{jj'}}\left({m_Y \over 589 GeV}\right)^3 \left( 1 - {\eta \, m_X \over m_Y}\right).
\ee An illustration of the inequality can be seen in Fig. \ref{fig:aboveineq}. Notice that for large couplings, i.e. $\alpha_{XXY}>0.1$ major portions
of the parameter space can be ruled out. The termination point ($\Gamma_Y \rightarrow 0$) is simply the $\eta m_X = m_Y$ point, giving us another
bound on $m_X$. By inspection, a coupling $\alpha_{XXY} = 0.1$ and $m_Y = 1.5 TeV$ requires $m_X \lesssim 600 GeV$.

\section{Conclusions}

The dynamical effects of resonant processes and finite particle widths play an important role in dark matter evolution in the early universe. The
Born approximation is seldom adequate because the non-relativistic velocity dependence of cross sections drives decoupling. Organizing
the calculation in terms of observable quantities gives new relations between the masses and widths of intermediate states that will be consistent with fixed
relic abundance.

Given that particle widths are generally calculable, our mass-v-width relations develop into mass-v-mass consistency relations between the dark matter
with a given relic density and the mass of an $s$-channel connector. Depending on the model, this produces a significant revision of a traditional
rule $\sigmav \sim 3 \times 10^{-26}\, cm^{3}/s$. The relation between $m_{X}$
and $m_{Y}$ depends on the way the width is calculated, but in a broad class of models permits an unlimited range of both masses. Our relations can be used to test candidates for dark matter in LHC-based experiments, while also eliminating much of the need
to re-compute relic evolution on a model-by-model basis.

\medskip

 \textbf{Acknowledgments:} We thank Yudi Santoso and KC Kong for helpful discussions. Research supported in part under DOE Grant Number DE-FG02-04ER14308.

\end{document}